\documentclass[journal]{IEEEtran}
\ifCLASSINFOpdf
   \usepackage[pdftex]{graphicx}
   \graphicspath{{./Images/pdf/}{./Images/jpeg/}{./Images/eps}}
   \DeclareGraphicsExtensions{.pdf,.jpeg,.png,.eps}
\else
   \usepackage[dvips]{graphicx}
   \graphicspath{{../eps/}}
   \DeclareGraphicsExtensions{.eps}
\fi

\usepackage{amsmath,amssymb,amsthm}
\usepackage{mathtools}
\usepackage{bbm}

\usepackage{svg}
\svgpath{{../svg/}} 
\usepackage{url}
\usepackage{multirow}
\usepackage[ruled, lined, linesnumbered, commentsnumbered, longend]{algorithm2e}

\usepackage{multicol}
\usepackage{subcaption}

\DeclareMathOperator*{\argmin}{argmin}

\newcommand{\integer}{\mathbb{Z}}
\newcommand\oprocendsymbol{\hbox{$\square$}}
\newcommand\oprocend{\relax\ifmmode\else\unskip\hfill\fi\oprocendsymbol}

\usepackage{amsmath,mathrsfs}
\DeclareMathOperator*{\minimize}{min}
\DeclareMathOperator{\subto}{subject~to}

\newtheorem{theorem}{Theorem}[section]

\newtheorem{remark}[theorem]{Remark}

\newenvironment{pfof}[1]{\vspace{1ex}\noindent{\itshape Proof of
    #1:}\hspace{0.5em}} {\hfill\oprocend\vspace{1ex}}
\newcommand{\RNum}[1]{\uppercase\expandafter{\romannumeral #1\relax}}

\usepackage{tikz}
\usetikzlibrary{spy}
\usetikzlibrary{backgrounds}
\usetikzlibrary{arrows}
\usetikzlibrary{arrows.meta}
\usetikzlibrary{decorations.pathmorphing}
\usetikzlibrary{decorations.markings}
\usetikzlibrary{snakes}
\usetikzlibrary{matrix}
\usetikzlibrary{calc}
\usetikzlibrary{patterns}
\usetikzlibrary{positioning}
\usetikzlibrary{shapes}
\usetikzlibrary{shapes.symbols}
\usetikzlibrary{shapes.geometric}
\usetikzlibrary{shadows.blur}
\usetikzlibrary{shapes.arrows}
\usetikzlibrary{shapes.multipart}
\usetikzlibrary{shapes.callouts}
\usetikzlibrary{shapes.misc}
\usetikzlibrary{fit}
\usepackage{makecell}

\tikzstyle{every node}=[font=\small]
\tikzstyle{every path}=[line width=0.8pt,line cap=round,line join=round]

      \tikzset{
  basic box/.style = {
    shape = rectangle,
    align = center,
    draw  = #1,
    fill  = #1!5,
    rounded corners}}

\usepackage{color}

\usepackage{soul}

\DeclareSymbolFont{bbold}{U}{bbold}{m}{n}
\DeclareSymbolFontAlphabet{\mathbbold}{bbold}

\newcommand{\real}{\mathbb{R}}

\begin{document}
%


\title{Data-Driven Fast Frequency Control using Inverter-Based Resources}

%
%
%

\author{Etinosa~Ekomwenrenren,~\IEEEmembership{Student Member,~IEEE,}
        John~W.~Simpson-Porco,~\IEEEmembership{Senior Member,~IEEE,}
        Evangelos~Farantatos,~\IEEEmembership{Senior Member,~IEEE,}
        Mahendra~Patel,~\IEEEmembership{Life~Fellow,~IEEE,}
        Aboutaleb~Haddadi,~\IEEEmembership{Senior~Member,~IEEE,}
        and~Lin~Zhu,~\IEEEmembership{Senior~Member,~IEEE}
\thanks{E. Ekomwenrenren is with the Department of Electrical and Computer Engineering, University of Waterloo, Waterloo, ON N3L 3G1 CA
(e-mail: etinosa.ekomwenrenren@uwaterloo.ca).}
\thanks{J.~W.~Simpson-Porco is with the Department of Electrical and Computer Engineering, University of Toronto, 10 King's College Road,
Toronto, ON, M5S 3G4, CA (e-mail: jwsimpson@ece.utoronto.ca)}\thanks{E. Farantatos, M. Patel, A. Haddadi and L. Zhu are with the Electric Power Research Institute, Palo Alto, CA 94304 USA
(e-mail: efarantatos@epri.com; mpatel@epri.com; ahaddadi@epri.com; lzhu@epri.com).}

\thanks{Funded under EPRI Project \#10009168: Wide-Area Hierarchical Frequency and Voltage Control for Next Generation Transmission Grids.}}

%
%

\markboth{Submitted for publication. This version: \today}%
{Shell \MakeLowercase{\textit{et al.}}: Bare Demo of IEEEtran.cls for IEEE Journals}
%



\maketitle

\begin{abstract}
To address the control challenges associated with the increasing share of inverter-connected renewable energy resources, this paper proposes a direct data-driven approach for fast frequency control in the bulk power system. The proposed control scheme partitions the power system into control areas, and leverages local dispatchable inverter-based resources to rapidly mitigate local power imbalances upon events. The controller design is based directly on historical measurement sequences, and does not require identification of a parametric power system model. Theoretical results are provided to support the approach. Simulation studies on a nonlinear three-area test system demonstrate that the controller provides fast and localized frequency control under several types of contingencies.
\end{abstract}


%
\IEEEpeerreviewmaketitle


\section{Introduction}

\IEEEPARstart{A} key objective in power system operations is the maintenance of a stable system frequency and the quick restoration of power balance \cite{anderson2008power}.  However, the increasing penetration of inverter-connected renewable energy resources (RESs) \cite{chakraborty2023hierarchical} is resulting in adverse effects on power system frequency regulation \cite{yan2018anatomy}. By replacing conventional synchronous generators, along with their synchronized rotational mass, the increasing proliferation of RESs reduces the system inertia, resulting in a faster rate of change of frequency (ROCOF) and lower frequency nadir during contingencies (i.e., a deeper drop in frequency). Furthermore, with some RESs not participating in primary frequency control\cite{dreidy2017inertia} while displacing synchronous generators which do, the aggregate effective primary control response in the system is reduced
\cite{ingleson2005tracking, vorobev2019deadbands}.
When combined with the net load variability these intermittent and variable RESs introduce, the upshot is larger and more frequency deviations, making it increasingly difficult for system operators to maintain  frequency within acceptable limits, such as the extreme variations noted by the California Independent System Operator in the so-called 'duck chart' \cite{osti_1226167}.

There has been extensive research into the negative dynamic effects of reduced inertia in the power system due to increased renewables, with suggested solutions such as virtual inertia emulation and services
\cite{stanojev2020mpc,poolla2017optimal}.
%
Equally important to consider is the problem of maintaining the average frequency close to nominal during normal operation, with regulation performance being quantified by regulatory authorities in Control Performance Standards (CPS) \cite{jaleeli1999nerc, yao2000agc}. For this, it is essential to consider the local system inertia, primary control response, and primary control deadband, as these have the greatest effect on the average frequency deviations \cite{vorobev2019deadbands}. This fosters the need for localized fast frequency control strategies, which take into cognizance local system model and parameter information. Traditionally, the Automatic Generation Control (AGC) system employs a centralized approach to maintain average frequency deviations within desired limits for each balancing authority area. This is achieved by generating control signals at a central control center. However, due to the sheer size of the balancing authority area, maintaining an accurate dynamic system model becomes an arduous task. Consequently, the AGC system relies on classical frequency bias constant methods \cite{baros2021examining}, which, while effective to some extent, limit its speed and utility for rapid frequency control. 

If accurate parametric models are available, then modern model-based controller design approaches can be successfully used to enable IBR participation in local fast frequency control. In  \cite{ekomwenrenren2021hierarchical} the authors developed and validated such a controller based on the principles of active disturbance estimation and rejection; as our work here builds directly on this, further details are deferred to Section \ref{Sec:Review}. This scheme 
and other model-based frequency control approaches, such as model predictive control
\cite{liu2019distributed, ademola2020frequency}
and robust optimal control \cite{bevrani2014robust,mavalizadeh2020decentralized} can provide good closed-loop control performance and reduce average frequency deviations. However, obtaining accurate parametric models may be prohibitively difficult in practice \cite{zholbaryssov2021safe}, ultimately limiting the performance of model-based designs. For example, simulations in \cite{ekomwenrenren2021hierarchical} show deterioration in control performance (e.g., post-
disturbance settling time and overshoot) when there are parametric mismatches between the true system model and the model used for design.

Data-driven or data-based controller design methods provide a promising alternative in this regard. Proposals to address the issue of power system frequency control using data-driven control can be broadly divided into two categories: indirect and direct. In an indirect approach,  historical data from the power system is used to explicitly identify a system model, and a controller is then designed based on that model (e.g., \cite{liu2019model, hidalgo2019frequency}). The indirect approach has the advantage of providing an explicit, interpretable model of the system, which can aid in understanding the particular frequency response dynamics. However, even selecting an appropriate parametric model to fit is a difficult trial-and-error process, particularly in modern power systems with diverse and quickly evolving components. Additionally, there is evidence that the intermediate identification step may lead to poorer closed-loop performance than recent direct approaches \cite{dorfler2022bridging}. 



In a direct data-driven or model-free approach, a frequency controller is designed directly based on recorded or online real-world data, without explicit identification of a system model. One broad approach in this category is adaptive dynamic programming or reinforcement learning \cite{chen2020model, yu2011stochastic, yan2020multi}.
Here, control actions are taken to maximize some form of cumulative reward. However, reinforcement learning approaches are limited by their sensitivity to hyper-parameter selection, the complex training process required to determine the weight coefficients of the trained agent, which in turn relies on a significant amount of historical sampled data \cite{coulson2019data,bucsoniu2018reinforcement}.

In contrast with reinforcement learning, a suite of alternative direct data-driven control approaches have recently appeared \cite{markovsky2008data, coulson2019data, de2019formulas, zhao2021data}, and derive from a branch of control theory called behavioral systems
\cite{willems2005note}.
These techniques allow for direct control while being sample efficient, and often come with rigorous performance guarantees. While the specific controllers differ between approaches (e.g., model-predictive \cite{coulson2019data}, linear-quadratic, \cite{de2019formulas} etc.), these approaches all rely on the so-called \emph{fundamental lemma} of behavioral systems, which states that a single recorded trajectory is sufficient to capture the underlying dynamic model of the system if the input signal is rich enough to excite all system modes \cite{willems2005note}. Our proposed controller is also based on this principle. 


\smallskip

\paragraph*{Contributions}

This paper provides direct data-driven controller designs which enable IBRs to participate in providing geographically localized fast frequency control. The key components in our approach are novel designs for \emph{data-driven disturbance estimators}: dynamic algorithms which provide online estimates of the net real power imbalance within a specified control area. Local IBRs are then quickly redispatched within their operating limits to eliminate the imbalance. 

In Section \ref{Sec:LocalController} we present two data-driven disturbance estimator designs. Both designs are based directly on recorded system data, and do not require a parametric system model. The two designs trade off between simplicity and robustness/performance. The first design, proposed in our preliminary work \cite{EE-JWSP-EF-MP-AH-LZ:22c}, uses a simple linear update law to estimate the disturbance, and requires tuning of only a single parameter. Extending \cite{EE-JWSP-EF-MP-AH-LZ:22c}, here we provide theoretical guarantees supporting this design. The first design serves as a stepping stone to our second approach, which is an optimization-based estimation procedure. The second design has a higher computational burden, requiring the solution of a convex optimization problem at each time-step, but (i) is less sensitive to noise in the recorded data, (ii) is less sensitive to strong nonlinearity in the system dynamics (e.g., governor deadbands), and (iii) shows superior performance in simulation studies. As the formulations are general, we outline specifically how these methods are applied to the frequency control problem under consideration. Compared, for instance, to the recent data-driven load-frequency controller proposed in \cite{zhao2021data}, we do not make the strong assumption that a measurement of net load demand is available; our approach is based only on direct measurements of area frequency and net power flow out of the control area. 

In Section \ref{Sec:CaseStudies} we extensively validate our designs via simulations on a detailed
nonlinear three-area power system. Several scenarios are examined, included load increases, heavy renewable penetration, generation trips, and three-phase faults. The tests demonstrate that our approach provides fast and effective frequency control for the bulk grid, and outperforms our recent model-based design \cite{ekomwenrenren2021hierarchical}.


\section{Review: Model-Based Fast Frequency Controller \cite{ekomwenrenren2021hierarchical}}
\label{Sec:Review}

We briefly review the key architectural aspects of the model-based hierarchical fast frequency controller developed and extensively tested in \cite{ekomwenrenren2021hierarchical}. The key criteria driving the design are a desire for (i) design simplicity, (ii) fast localized response to achieve frequency regulation, and (iii) localized use of system data and measurements to minimize latency and maximize data privacy.

\begin{figure}[ht!]
\begin{center}
    \begin{tikzpicture}[auto, scale = 0.6, node distance=2cm,>=latex', every node/.style={scale=0.9},
      blacknode/.style={shape=circle, draw=black, line width=2},
  bluenode/.style={shape=circle, draw=blue, line width=2},
  greennode/.style={shape=circle, draw=green, line width=2},
  rednode/.style={shape=circle, draw=red, line width=2},
  dotted_block/.style={draw=blue!40!white, line width=1pt, dash pattern=on 1pt off 4pt on 6pt off 4pt,
            inner ysep=6mm,inner xsep=4mm, rectangle, rounded corners}
    ]
      \tikzstyle{anch} = [coordinate]
      \tikzstyle{block} = [draw, fill=white, rectangle, 
      minimum height=3em, minimum width=6em]
      \tikzstyle{bigblock} = [draw, fill=white, rectangle, 
      minimum height=5em, minimum width=25em]
      \tikzstyle{wideblock} = [draw, fill=white, rectangle, 
      minimum height=3em, minimum width=9em]
      \tikzstyle{hold} = [draw, fill=white, rectangle, 
      minimum height=1.5em, minimum width=2em]
      \tikzstyle{delayblock} = [draw, fill=white, rectangle, 
      minimum height=1em, minimum width=0.7em, inner sep=2pt, outer sep=0pt]
      \tikzstyle{sum} = [draw, fill=white, circle, minimum size=0.2cm, node distance=1cm, inner sep=0pt, outer sep=0pt]
      \tikzstyle{mynode} = [draw, thick, fill=white, circle]
      \tikzstyle{input} = [coordinate]
      \tikzstyle{cblue}=[circle, draw, thick, fill=white, scale=1]
      \tikzstyle{output} = [coordinate]

      \node[block] (power_system)  {Local Control Area};
      \node[block, below of=power_system,fill=blue!20, node distance=3cm, xshift=1.25cm, name=] (estimator) {\makecell[c]{Disturbance \\Estimator}};
      
      
      \node[hold, left of=estimator,fill=blue!20, node distance=3cm] (allocator) {\makecell[c]{Power \\Allocator}};
      \node [dotted_block, fit = (estimator) (allocator),inner ysep=4mm, inner xsep=5mm,xshift=0mm,yshift=-2mm] (controller) {};
      \node at (controller.south) [above, inner sep=1mm,] {LCA Controller};

      \node[hold, left of=power_system, node distance=3cm, yshift=-0.18cm, name=IBRs] {IBRs};

      \node[anch, above of=estimator, node distance=1.5cm, xshift=0.65cm] (sum2) {};
      \node[anch, above of=estimator, node distance=1cm, xshift=-0.65cm] (sum3) {};
      \node[anch, right of=estimator, node distance=1.7cm] (sum4) {};


	  \node[delayblock, left of=sum2, node distance=1.5cm] (delay2) {\footnotesize {$z^{-\tau_{\mathrm{m}}}$}};
	  \node[delayblock, left of=sum3, node distance=1.5cm] (delay3) {\footnotesize {$z^{-\tau_{\mathrm{m}}}$}};	  
	  \node[delayblock, above of=sum4, node distance=1cm] (delay4) {\footnotesize {$z^{-\tau_{\mathrm{m}}}$}};

      \node[input, above of=power_system, node distance=1.2cm, name=load] {};      
      \node[input, left of=power_system, node distance=4.5cm, name=Ptie] {};   
      \node[output, right of=power_system, node distance=3.5cm, name=w] {};   
      
      \node[anch, left of = IBRs, node distance=1.5cm] (tmp1) {};

      \node[delayblock, below of=tmp1, node distance=1.5cm] (delay_ibr) {\footnotesize {$z^{-\tau_{\mathrm{c}}}$}};

      \draw [thick,-latex] (IBRs) -- node[name=p_ibr,pos=0.3] {} ([yshift=-0.3cm]power_system.west);
      \draw[thick,dashed,-latex] (estimator) -- node[above] {$\Delta \widehat{P}_{\rm u}$} (allocator);

      \draw [thick,-latex] ([yshift=0.4cm]Ptie) -- node[above, pos=0.67, name = p_tie] {$\Delta P_{\mathrm{tie}}$} ([yshift=0.4cm]power_system.west);
      
      \draw [thick, -latex] (load) -- node[pos=0, right] {$\Delta P_{\mathrm{u}}$} (power_system);
      \draw [thick,-latex] (power_system.east) -- node[pos=0.75, name=freq] {$\Delta \omega$} (w);

      \draw [thick,dashed,-latex] (freq) -- (delay4);
      \draw [thick,dashed,-latex] (delay4) -- (sum4) -- (estimator);

	  \draw [thick,dashed,-latex] (allocator.west) -| node[pos = 0.2, left, xshift=-0.2cm] {} (delay_ibr) |- node[pos=0.64, left, xshift=-0.2cm] {$\Delta P_{\mathrm{ibr}}^{\rm set}$} (IBRs) ;
	  
	  \draw [thick,dashed, -latex] ([xshift=7cm]p_tie) |- (delay2);
	  \draw [thick,dashed, -latex] (delay2) -- (sum2) -- ([xshift=1cm]estimator.north);
	  
	  \draw [thick,dashed, -latex] (p_ibr) |- node[pos=0.3, left] {$\Delta P_{\mathrm{ibr}}$} (delay3);
	  \draw [thick,dashed, -latex] (delay3) -- (sum3) -- ([xshift=-1cm]estimator.north);

      \end{tikzpicture}
      \end{center}
      \caption{Block diagram of area control structure for each LCA. Dashed lines denote sampled signals.}
      \label{Fig:Block}
\end{figure}
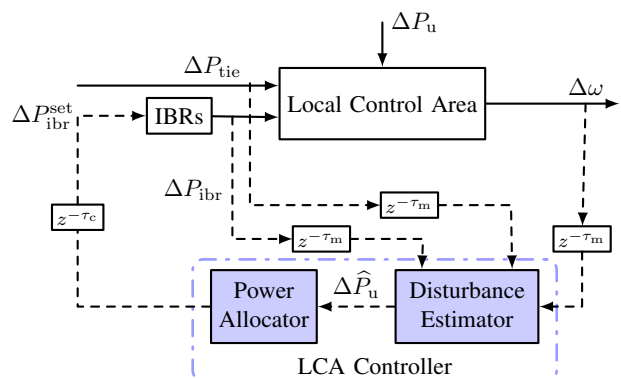

In the scheme of \cite{ekomwenrenren2021hierarchical}, the grid is divided into small local control areas (LCAs). Within each LCA, a \emph{disturbance estimator} processes frequency and area tie power flow measurements $\Delta \omega$ and $\Delta P_{\rm tie}$ (incorporating estimated delays) in order to detect frequency events. The estimator generates a real-time estimate $\Delta \widehat{P}_{\rm u}$ of the net unmeasured active power imbalance $\Delta P_{\rm u}$ within the LCA, and an allocation mechanism optimally redispatches local IBRs to correct the imbalance; see \cite[Section III-A]{ekomwenrenren2021hierarchical} for further details on the power allocator. A block diagram of this LCA controller is shown in Figure \ref{Fig:Block}.
In situations where local resources are insufficient, a higher-level coordinating controller facilitates the provision of additional power support from neighboring LCAs; as this higher-level coordinating controller is not our focus in this article, we refer to \cite[Section III-B]{ekomwenrenren2021hierarchical} for further details. 

The disturbance estimator within the LCA controller is the core design component of the approach. The estimator is a Luenberger-type state observer, and its design requires a parametric model describing the LCA frequency dynamics. For practical reasons, a simple model with few parameters is strongly preferred, and the designs in \cite[Section II]{ekomwenrenren2021hierarchical}, used the second-order \emph{frequency response model}
\begin{equation}\label{Eq:SFR}
\resizebox{0.9025\linewidth}{!}{$
\begin{aligned}
    2H\Delta{\dot{\omega}} &= - \tfrac{1}{R_{\mathrm I}}\Delta \omega + \Delta P_{\mathrm{m}} - \Delta P_{\mathrm{u}} - \Delta P_{\mathrm{tie}} + \Delta P_{\mathrm{ibr}}\\
    T_{\rm R}\Delta{\dot{P}_{\mathrm{m}}} &= - \Delta P_{\mathrm{m}} - R_{\rm g}^{-1}(\Delta{\omega} + T_{\mathrm{R}}F_{\mathrm{H}}\Delta{\dot{\omega}})
\end{aligned}
$}
\end{equation}
which considers local aggregated parameters, such as total LCA inertia $H$, total IBR and generator primary control gains $R_{\rm I}$ and $R_{\rm g}$,  aggregated turbine-governor time constant $T_{\rm R}$, and aggregated  high-pressure turbine fraction $F_{\rm H}$.


\section{A Data-Driven Control Approach for Area-Based Fast Frequency Control}
\label{Sec:LocalController}

The model-based design of Section \ref{Sec:Review} requires an explicit and accurate model of the frequency dynamics of each LCA. In practice, this requirement poses at least two major challenges. First, an appropriate class of parametric models must be selected; this step balances simplicity vs. accuracy, and will become increasingly difficult as RESs with black-box power electronic controls proliferate. Second, the parameters of the model must be selected or fit; this procedure itself is challenging, with associated bias-variance trade-offs \cite{LL-TC-BM:20}. 

To address these issues, in this section we develop two direct data-driven design approaches to supplant the model-based design approach described in Section \ref{Sec:Review}. In essence, the idea is to replace the crude parametric LCA model \eqref{Eq:SFR} with a non-parametric model based on time-series data collected from the system. This time-series data is directly used to design a disturbance estimation scheme, without passing through an explicit system identification step. 




Section \ref{Sec:LTIDisturbanceEstimator} describes our first data-driven disturbance estimation approach, which fuses ideas from linear estimator design and behavioral systems theory. The resulting disturbance estimator is described by a linear update rule, and requires tuning of only one scalar gain. To improve robustness to grid nonlinearities and inexact data collection procedures, our second design approach in Section \ref{Sec:RobustDisturbanceEstimator} extends this linear estimation procedure with an optimization-based estimation procedure. Finally, in Section \ref{Sec:FastFreqControl} we describe how these general estimation ideas are adapted for the particulars of power system frequency control and integrated into the hierarchical control framework outlined in Section \ref{Sec:Review}.



\subsection{Background on Data-Driven System Representation}
\label{Sec:Background}

Consider the controllable finite-dimensional discrete-time linear time-invariant (LTI) model
\begin{equation}\label{Eq:LTI}
\begin{aligned}
x(t+1) &= Ax(t) + Bu(t) + B_d d(t)\\
y(t) &= Cx(t) + Du(t)
\end{aligned}
\end{equation}
with time $t \in \integer_{\geq 1}$, state $x(t) \in \real^n$, control input $u(t) \in \real^m$, disturbance input $d(t) \in \real^{q}$, and measured output $y(t) \in \real^{p}$. We assume the matrices $(A,B,B_d,C,D)$ of \eqref{Eq:LTI} are \emph{unknown}, and hence the model \eqref{Eq:LTI} cannot be directly used for simulation, analysis, and feedback design purposes. \emph{Behavioral systems theory} provides a set of tools for constructing a data-based representation of the dynamic system \eqref{Eq:LTI} using input and output measurements. We refer to \cite{IM-FD:21b} for a recent survey, and mention only the essential concepts here.

As notation, if $(z(1),z(2),z(3),\ldots)$ is a $\real^m$-valued signal defined for positive time, we write $z \in (\real^m)^{\integer_{\geq 1}}$. The starting point is to diminish the role of the state, and consider all possible input-output sequences $(u(t),d(t),y(t))$ which are compatible with \eqref{Eq:LTI}, called the  \emph{behaviour}:
\begin{equation}\label{Eq:Behaviour}
\begin{aligned}
\mathscr{B} &= \left\{(u,d,y) \in (\real^{m+q+p})^{\integer_{\geq 1}}\,:\,\exists \,x\in(\real^{n})^{\integer_{\geq 1}}\,\,\text{s.t.}\right.\\
& \qquad\left.\sigma x = Ax + Bu + B_dd,\,\, y = Cx+Du\right\},
\end{aligned}
\end{equation}
where $(\sigma x)(t) = x(t+1)$ is the shift operation. The behaviour \eqref{Eq:Behaviour} describes the system \eqref{Eq:LTI} as a subspace of the vector space of all possible input-output signals, and \eqref{Eq:LTI} is a \emph{state-space representation} of $\mathscr{B}$. The \emph{order} of the system, denoted by $n(\mathscr{B})$, is the smallest possible state dimension of the representation \eqref{Eq:LTI}. Given a representation of minimal order, the \emph{lag} of $\mathscr{B}$, denoted by $\ell(\mathscr{B})$ is the smallest integer $\ell$ such that the matrix $\mathcal{O}_{\ell} = \mathrm{col}(C,CA,\ldots,CA^{\ell-1})$ has rank $n(\mathscr{B})$. 

Let $\mathscr{B}_{T}$ denote the restriction of the behaviour to trajectories of finite length $T \in \integer_{\geq 1}$, i.e., input-output sequences of length $T$. Suppose that we have collected $T$-samples of \emph{input-output data} $w^{\rm d} = (u^{\rm d},d^{\rm d},y^{\rm d}) \in \mathscr{B}_{T}$ from the system. This data may be directly used to create a non-parametric representation of the model \eqref{Eq:Behaviour}. To do this, let $L \leq T$ be a positive integer, and organize the data into the \emph{Hankel matrix of depth $L$}, given as
\[
\mathscr{H}_{L}(u^{\rm d}) = \begin{bmatrix}u^{\rm d}(1) & \cdots & u^{\rm d}(T-L+1)\\
\vdots & \ddots & \vdots\\
u^{\rm d}(L) & \cdots & u^{\rm d}(T)
\end{bmatrix} \in \real^{mL \times (T-L+1)},
\]
with analogous definitions for $\mathscr{H}_{L}(d^{\rm d})$ and $\mathscr{H}_{L}(y^{\rm d})$. The input data $(u^{\rm d},d^{\rm d})$ is said to be \emph{persistently exciting of order $L$} if $\mathrm{col}(\mathscr{H}_{L}(u^{\rm d}),\mathscr{H}_{L}(d^{\rm d}))$ has full row rank; this captures the idea that the inputs are sufficiently rich and sufficiently long. The \emph{Fundamental Lemma} \cite{willems2005note} states that if the input data is persistently exciting of order $L + n(\mathscr{B})$, then \emph{any possible} length $L$ input-output sequence $(u,d,y) \in (\real^{(m+q+p)})^{L}$ can be expressed as
\begin{equation}\label{Eq:FundLemma}
\begin{bmatrix}
\mathscr{H}_{L}(u^{\rm d})\\
\mathscr{H}_{L}(d^{\rm d})\\
\mathscr{H}_{L}(y^{\rm d})
\end{bmatrix}g = \begin{bmatrix}u \\ d \\ y\end{bmatrix}
\end{equation}
for some vector $g \in \real^{T-L+1}$. The linear equation \eqref{Eq:FundLemma} is a \emph{data-based representation} of the system \eqref{Eq:LTI}, and can be leveraged for prediction and control \cite{IM-FD:21b}.

\subsection{Design \#1: Linear Data-Driven Disturbance Estimator}
\label{Sec:LTIDisturbanceEstimator}

We now consider \eqref{Eq:LTI} as a model for each LCA. 
We assume that $d(t)$ is a constant unknown disturbance signal, which for us will model mismatch between generation and load. In our context, $x(t)$ would consist of states of generators, converters, loads, and associated control systems, $u(t)$ would be commands to IBRs, and $y(t)$ would be available measurements such as frequency deviation. Since \eqref{Eq:LTI} would represent the system \emph{including} the action of primary controllers, the model \eqref{Eq:LTI} will be assumed to be internally exponentially stable, i.e., $A$ will have eigenvalues within the unit circle.

The design goal is to produce a real-time estimate $\hat{d}(t)$ of the unknown disturbance $d(t)$. Our proposed estimator design consists of two steps:
\begin{enumerate}
\item[(i)] a \emph{data-driven forward prediction} $\hat{y}(t)$ of the output $y(t)$;
\item[(ii)] a \emph{linear update rule} for $\hat{d}(t)$ using $\hat{y}(t)$ and the true system measurement $y(t)$.
\end{enumerate}

To generate a prediction of the output for time $t$, we will leverage \eqref{Eq:FundLemma}, and assume that historical data $(u^{\rm d},d^{\rm d},y^{\rm d})$ is available. Model-based prediction using, e.g., \eqref{Eq:LTI} would require the specification of an initial condition. In the data-driven setting, the initial condition is implicitly defined by using recent online samples of input and output data \cite{markovsky2008data}. Let $T_{\rm ini} \geq \ell(\mathscr{B})$ be the length of the initialization data, and define the vectors
\begin{equation}\label{Eq:DataCollection}
\begin{aligned}
u_{\rm ini} &= \mathrm{col}(u(t-T_{\rm ini}),\ldots,u(t-1))\\
\hat{d}_{\rm ini} &= \mathrm{col}(\hat{d}(t-T_{\rm ini}),\ldots,\hat{d}(t-1))\\
\hat{y}_{\rm ini} &= \mathrm{col}(\hat{y}(t-T_{\rm ini}),\ldots,\hat{y}(t-1)).
\end{aligned}
\end{equation}
Note that $\hat{d}_{\rm ini}$ and $\hat{y}_{\rm ini}$ are formed based on our past \emph{estimates} of the  disturbance and output. In \eqref{Eq:FundLemma}, we consider trajectories of length $L = T_{\rm ini} + 1$. We partition $u,d,y$ in \eqref{Eq:FundLemma} as
\[
u = \begin{bmatrix}u_{\rm ini} \\ u(t)\end{bmatrix}, \quad \hat{d} = \begin{bmatrix}d_{\rm ini} \\ \hat{d}(t)\end{bmatrix}, \quad y = \begin{bmatrix}y_{\rm ini} \\ \hat{y}(t)\end{bmatrix},
\]
and correspondingly partitioning the rows of the Hankel matrices in the same fashion as
\[
\mathscr{H}_{L}(u^{\rm d}) = \begin{bmatrix}U_{\rm ini} \\ U_f\end{bmatrix},\,\,\mathscr{H}_{L}(d^{\rm d}) = \begin{bmatrix}D_{\rm ini} \\ D_f\end{bmatrix},\,\, \mathscr{H}_{L}(y^{\rm d}) = \begin{bmatrix}Y_{\rm ini} \\ Y_f\end{bmatrix}.
\]
With these choices, \eqref{Eq:FundLemma} can be re-expressed as 
\begin{equation}\label{Eq:DataDrivenSimulation1}
\mathscr{H}_{\rm red}g := \begin{bmatrix}
U_{\rm p}\\
D_{\rm p}\\
Y_{\rm p}\\
U_{f}\\
D_{f}
\end{bmatrix}g = \begin{bmatrix}
u_{\rm ini}\\
\hat{d}_{\rm ini}\\
\hat{y}_{\rm ini}\\
u(t)\\
\hat{d}(t)
\end{bmatrix}, \qquad \hat{y}(t) = Y_fg.
\end{equation}
The first set of equations is solved for the unknown $g$, and the prediction $\hat{y}(t) = Y_f g$ is immediately obtained. If the underlying data-generating system is LTI and the collected data are exact, the Fundamental Lemma guarantees that \eqref{Eq:DataDrivenSimulation1} is consistent and the computed response matches the system's response exactly, provided $T_{\rm ini} \geq \ell(\mathscr{B})$ \cite{willems2005note}. 

With the output estimate generated, the disturbance estimate is now updated according to the feedback rule
\[
\hat{d}(t+1) = \hat{d}(t) - \varepsilon L(\hat{y}(t) - y(t)),
\]
where $L \in \real^{q \times p}$ is the estimation gain and $\varepsilon \in (0,1)$ is a tunable parameter which controls the rate of adjustment. Putting everything together, we can compactly express the overall disturbance estimator as
\begin{subequations}\label{Eq:DistEst1}
\begin{align}
\label{Eq:DistEst1-a}
\hat{y}(t) &= \mathcal{P} \cdot \mathrm{col}(u_{\rm ini},
\hat{d}_{\rm ini},
\hat{y}_{\rm ini},
u(t),
\hat{d}(t))\\
\label{Eq:DistEst1-b}
\hat{d}(t+1) &= \hat{d}(t) - \varepsilon L(\hat{y}(t) - y(t))
\end{align}
\end{subequations}
where $\mathcal{P} = Y_f\mathscr{H}_{\rm red}^{\dagger}$ is the \emph{prediction matrix} and $\mathscr{H}_{\rm red}^{\dagger}$ denotes the pseudoinverse of $\mathscr{H}_{\rm red}$. As $\mathcal{P}$ depends only on historical data, it can be computed once and stored, and thus implementing \eqref{Eq:DistEst1} simply amounts to matrix-vector multiplication.

The final issue to address concerns the tuning of the estimator gain $L$ and parameter $\varepsilon$ in \eqref{Eq:DistEst1}. Our tuning recommendation is $L = G(1)^{\dagger}$, where $G(1) = C(zI_n-A)^{-1}B_d|_{z=1}$ is the \emph{DC gain} of the system \eqref{Eq:LTI} from input $d$ to output $y$. This selection will be justified in our theory to follow, and the matrix $G(1)$ can be obtained directly from the same historical data used to construct $\mathcal{P}$ in \eqref{Eq:DistEst1}; see \cite[Thm. 4.1]{GB-MV-JC-ED:21} for details on that construction. We can now give a theoretical result concerning convergence of the disturbance estimator \eqref{Eq:DistEst1}.

\smallskip



\begin{theorem}[\bf Data-Driven Disturbance Estimator]\label{Thm:DDDE}
Consider the disturbance estimator \eqref{Eq:DistEst1} for the system \eqref{Eq:LTI} under all previous assumptions. Assume further that $G(1) = C(I_n-A)^{-1}B_d$ has full column rank, and set the estimator gain as $L = G(1)^{\dagger}$. Then there exists $\varepsilon^{\star} > 0$ such that for all $\varepsilon \in (0,\varepsilon^{\star})$, $\hat{d}(t) \to d(t)$ exponentially as $t \to \infty$.
\end{theorem}

The disturbance estimator \eqref{Eq:DistEst1} provides a \emph{completely model-free} solution to disturbance estimation problem; the only required tuning is the single scalar parameter $\varepsilon \in (0,1)$. An implication of Theorem \ref{Thm:DDDE} is that one may tune the estimator \eqref{Eq:DistEst1} by starting $\varepsilon$ small and slowly increasing it; the proof can be found in Appendix \ref{Append:A}.  

\begin{remark}[Singular Value Thresholding]
In practice, the system generating the data which is used to build $\mathscr{H}_{\rm red}$ in \eqref{Eq:DistEst1} may contain nonlinearity, and the measurements will be corrupted by measurement noise; this will be the case in our subsequent case studies. Both of these effects will compromise performance of the design \eqref{Eq:DistEst1}. It has however been observed that low-rank approximations of Hankel matrices reduce the effects of noise in data-driven control, and enhance generalization \cite{coulson2019data}. In implementation, we compute the singular value decomposition of $\mathscr{H}_{\rm red}$ and retain only the dominant singular values and vectors, to obtain a low-rank approximation $\tilde{\mathscr{H}}_{\rm red}$ \cite{golub2013matrix}. We then use $\mathcal{P} = Y_f\tilde{\mathscr{H}}_{\rm red}^{\dagger}$ in \eqref{Eq:DistEst1}, which, empirically, greatly increases the robustness of the approach. \hfill \oprocend
\end{remark}

\subsection{Design \#2: Optimization-Based Data-Driven Disturbance Estimator}
\label{Sec:RobustDisturbanceEstimator}

The advantage of \eqref{Eq:DistEst1} is simplicity, as it involves only linear update rules at each time step. We now outline a more flexible optimization-based disturbance estimation procedure which can achieve improved performance at the cost of higher implementation complexity. The key idea is to formulate the disturbance estimation problem as a regularized optimization problem. In particular, the use of regularization affords us more flexibility to select a better model class in terms of behaviour and complexity to better capture the dynamics of the true underlying system \cite{IM-FD:21b}.

To begin, consider the previous development leading up to equation \eqref{Eq:DataDrivenSimulation1}. Even if the system of equations  \eqref{Eq:DataDrivenSimulation1} is consistent, it will generally have infinitely many solutions \cite{markovsky2008data,dorfler2022bridging}. The prediction matrix $\mathcal{P}$ in \eqref{Eq:DistEst1} is given by $\mathcal{P} = Y_f\mathscr{H}_{\rm red}^{\dagger}$, and corresponds precisely to taking the \emph{least squares} solution of the first equation in \eqref{Eq:DataDrivenSimulation1} as
\[
\begin{aligned}
    g^{\star} = \argmin_{g} &\quad \|g\|_2^2 &&\\
    \subto &\quad \mathscr{H}_{\rm red}g = \mathrm{col}(u_{\rm ini},
\hat{d}_{\rm ini},
\hat{y}_{\rm ini},
u(t),
\hat{d}(t))
\end{aligned}
\]
and then substituting to obtain $\hat{y}(t) = Y_{f}g^{\star}$. When using noisy data from a non-LTI data-generating system, it is advantageous to robustify this least-squares problem by adding regularization \cite{dorfler2022bridging}. To this end, for the equation $\mathcal{H}_{\rm red}g = \xi$, we have
\[
g = \mathscr{H}_{\rm red}^{\dagger}\xi \qquad \Longleftrightarrow \qquad (I-\mathscr{H}_{\rm red}^{\dagger}\mathscr{H}_{\rm red})g = 0.
\]
Thus, with $\mathcal{Q} = \mathscr{H}_{\rm red}^{\dagger}\mathscr{H}_{\rm red}$, a least squares solution for $g$ also arises from minimizing the objective function $\|(I -\mathcal{Q})g\|_2^{2}$ subject to the linear constraint $\mathscr{H}_{\rm red}g = \xi$. 

Our disturbance estimation approach is now to \emph{intentionally bias} this least squares solution, by introducing additional objective functions quantifying the prediction error along with regularization on $g$. This intentional biasing exploits the bias-variance trade-off from system identification \cite{LL-TC-BM:20}, leading to reduced overfitting in the estimation procedure. With the same notation and set-up as in Section \ref{Sec:LTIDisturbanceEstimator}, at time $t$ we solve the convex optimization problem
\begin{equation}\label{RegularizedDDEstimator}
    \begin{aligned}
    \displaystyle{\minimize_{\hat{d}(t),\hat{y}(t),g} \,}&\quad \|y(t)-\hat{y}(t)\|_2^2 + \lambda_1\|(I -\mathcal{Q})g\|_2^{2}+\lambda_2\|g\|_2\\
    \text{s.t.} & \qquad \begin{bmatrix}
U_{\rm p}\\
D_{\rm p}\\
Y_{\rm p}\\
U_{f}\\
D_{f}\\
Y_f
\end{bmatrix}g = \begin{bmatrix}
u_{\rm ini}\\
\hat{d}_{\rm ini}\\
y_{\rm ini}\\
u(t)\\
\hat{d}(t)\\
\hat{y}(t)
\end{bmatrix},
    \end{aligned}
\end{equation}
where $\lambda_1, \lambda_2 \geq 0$ are tuning parameters. The problem \eqref{RegularizedDDEstimator} combines the prediction and estimation steps from \eqref{Eq:DistEst1} into one formulation, jointly generating the output prediction $\hat{y}(t)$ and the disturbance estimate $\hat{d}(t)$. The first objective function term attempts to match the prediction $\hat{y}(t)$ to the measured output $y(t)$. Increasing $\lambda_1$ encourages a least-squares solution for $g$, similar to that used in \eqref{Eq:DistEst1}, while increasing $\lambda_2$ regularizes the solution; this reduces overfitting \cite{LL-TC-BM:20} and improves estimation robustness for noisy measurements and non-LTI dynamics. While a theoretical estimation error analysis for \eqref{RegularizedDDEstimator} is outside the scope of this article, the approach is strongly justified by recent advances in regularized data-driven control \cite{IM-FD:21b}, and performance will be extensively tested in Section \ref{Sec:CaseStudies}.

\subsection{Specialization to Area-Based Fast Frequency Control using Inverter-Based Resources}\label{Sec:FastFreqControl}

We now describe the adaptation of our general data-driven disturbance estimation methods to the fast frequency control architecture described in Section \ref{Sec:Review}. Consider a large interconnected power system which is divided into several small LCAs. Each LCA has local IBR's that can be re-dispatched by the operator, subject to their real-time capacity limits. Since each LCA is geographically small, the effect of a power imbalance within the LCA on the frequency is approximately independent of the specific nodal location of the imbalance within the LCA. Therefore, it is assumed that power disturbances and generation are aggregate, and effectively lumped at a single bus. Put differently, disturbance and control signals enter through the same channel, and thus $B = B_{d} \in \real^{n \times 1}$ in \eqref{Eq:LTI}. 

The following selections are made for inputs and outputs: the measurement $y(t) = \Delta \omega(t) \in \real$ is a single local measurement of frequency deviation, and the disturbance $d(t) = \Delta P_{\rm u} \in \real$ models aggregate unmeasured generation-load imbalance within the LCA. The input $u(t)$ to the system consists of the measured tie-line flow $\Delta P_{\mathrm{tie}}(t)$ out of the LCA, as well as the sum of all IBR power set-points $\Delta P_{\mathrm{ibr}}(t)$. 

Historical data must be used to build the Hankel matrices used in both estimators. As the control and disturbance channels are lumped, during the collection of historical data, the sum of IBR set-point changes, exogenous load/generation changes, and inter-LCA tie-line flow changes must be recorded. Further discussion of options for data collection is deferred to Section \ref{Sec:OfflineData}.

As a result of the above, the estimator \eqref{Eq:DistEst1} simplifies to
\begin{subequations}\label{Eq:DistEst1-LCA}
\begin{align}
\Delta\hat{f}(t) &= \mathcal{P} \cdot \mathrm{col}(\Delta v_{\rm ini},
\Delta \hat{f}_{\rm ini},
\Delta v(t))\\
\Delta \hat{P}_{\rm u}(t+1) &= \Delta \hat{P}_{\rm u}(t) - \varepsilon L(\Delta \hat{f}(t) - \Delta f(t)),
\end{align}
\end{subequations}
where $\Delta v = \Delta P_{\rm ibr} - \Delta P_{\rm tie} - \Delta P_{\rm u}$, is the aggregated input, $\mathcal{P} = Y_f\left[\begin{smallmatrix}U_p;\,Y_p;\,U_f\end{smallmatrix}\right]^{\dagger}$, is the prediction matrix, and $L \in \real$ is now a scalar. Analogously, the optimization-based estimator \eqref{RegularizedDDEstimator} becomes
\begin{equation}\label{Eq:RegularizedDDEstimator-LCA}
    \begin{aligned}
    \displaystyle{\minimize_{} \,}&\,\, \|\Delta f(t)-\Delta \hat{f}(t)\|_2^2 + \lambda_1\|(I -\mathcal{Q})g\|_2^{2}+\lambda_2\|g\|_2\\
    \text{s.t.} & \,\, \begin{bmatrix}
U_{\rm p}\\
Y_{\rm p}\\
U_{f}\\
Y_f
\end{bmatrix}g = \begin{bmatrix}
\Delta P_{\rm ibr,ini} - \Delta P_{\rm tie,ini} - \Delta \hat{P}_{\rm u, ini}\\
\Delta f_{\rm ini}\\
\Delta P_{\rm ibr}(t) - \Delta P_{\rm tie}(t) -\Delta \hat{P}_{\rm u}(t)\\
\Delta \hat{f}(t),
\end{bmatrix},
    \end{aligned}
\end{equation}
The imbalance estimate $\Delta \hat{P}_{\rm u}(t)$ from either method is then used to redispatch the local IBRs in the LCA via the optimal power allocation algorithm presented in \cite{ekomwenrenren2021hierarchical}.

\section{Simulation Studies}
\label{Sec:CaseStudies}



\begin{figure}[ht!]
\centering
\includegraphics[width =1\columnwidth]{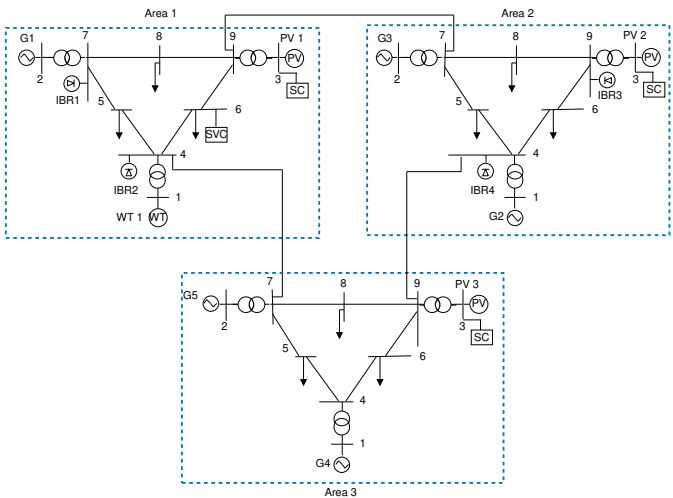}
\setlength{\belowcaptionskip}{-10pt}
\caption{Three-LCA test system.}
\label{fig:ThreeLCASystem}
\end{figure}

We validate our designs by applying them to the three-area nonlinear test system illustrated in Figure \ref{fig:ThreeLCASystem}. Each LCA of the test system is based on the the IEEE 3-machine 9-bus system given in \cite{anderson2008power}, with the interconnection parameters and active power dispatch info similar to \cite{ekomwenrenren2021hierarchical}. In the modified test model, two synchronous generators (SGs) in area one have been replaced with a photovoltaic (PV) array and a Wind (WT) plant. Similarly, one SG each in areas two and three are replaced with a PV farm. The PV array and wind turbine plant are simplified models represented by non-dispatchable converter-based units, which are parameterized using wind power and solar irradiance data from \cite{Godahewa2020, sengupta2010oahu}. To facilitate frequency control, two dispatchable IBRs have been added in each LCA. In addition, static var compensators (SVCs) and synchronous condensers have been added to areas 1 and 2/3 to support the voltage. All SGs and dispatchable IBRs in the system are set to have a 5\% speed droop curve on their respective bases, with a 36 mHz primary control deadband. The pre-disturbance generation/demand in the system is approximately 800 MW. 

\subsection{Offline Data Collection and Controller Tuning}
\label{Sec:OfflineData}

As described in Section \ref{Sec:LocalController}, our estimators require a library of historical data generated from a \emph{persistently exciting input} that must be collected \emph{before} the online implementation of the control. Examples of common persistently exciting inputs from the literature include, pseudo random binary sequence, autoregressive moving average sequence, sum of sinusoids, and white noise \cite{soderstrom1989system,ljung1999system}. Among these, white-noise derivatives are most commonly used in power system identification studies, such as measured ambient power fluctuations \cite{wies2003use}, and low-power injected probing signals such as the low-level pseudo-random noise (LLPRN) in \cite{zhou2006initial} and the band-limited white noise in \cite{huang2021decentralized}. In terms of what sources should be actuated for this data collection, there are several theoretically-equivalent options for the purposes of this work, including
\begin{enumerate}
\item[(i)] apply low-power probing modulations to IBRs during calm system conditions (i.e., during times of minimal unmeasured generation/load changes), and meter the resulting frequency and tie-line power changes,
\item[(ii)] hold IBR set-points constant, record ambient load power consumption changes (or injected pseudo random white noise that mimics such changes), and meter the resulting frequency and tie-line power changes,
\end{enumerate}
or obvious variations/combinations of these. Still other possibilities, such as using historical load estimates as proxy data, are of interest, but are outside of our scope in this study. In our testing to follow, we pursue option (i); we refer the reader to Remark III.1 in \cite{ekomwenrenren2021hierarchical} for a discussion on the feasibility and market incentives that makes this choice viable. 

\begin{figure}[ht!]
\centering
\includegraphics[width=1\columnwidth]{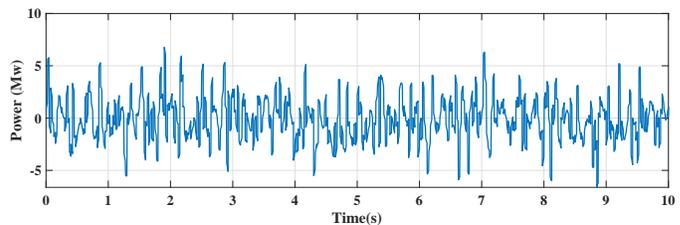}
\caption{Persistently exciting IBR set-point change for data collection phase.}
\label{fig:ExcitationSignal}
\end{figure}

We now turn to the design of the low-power probing signal for IBR set-point changes. In this work, we modeled our probing injection signal after the LLPRN in \cite{zhou2006initial}, where we have combined a sum of sinusoids and band-limited white noise. Each IBR within each LCA is commanded with the set-point changes shown in Figure \ref{fig:ExcitationSignal}, given by
\begin{equation}\label{Eq:IBRPerturb}
\Delta P_{\rm ibr}(t) = \sin(12\pi t) + w(t) \qquad (\text{in MW}).
\end{equation}
The signal consists of a sinusoidal perturbation of 1 MW ($1.76 \times 10^{-3}$ p.u.), with band-limited white noise $w(t)$ with noise power of $\approx$ $0.2 \times 10^{-3}$ p.u.\footnote{Further investigation into excitation signal design is outside our scope, but see \cite{de2021designing,van2021beyond} for recent theoretical results.}  


While we stress that the choice of probing signal is not unique, with our choice of signal, we are able to utilize the aggregated power input $u = \Delta P_{\mathrm{ibr}} - \Delta P_{\mathrm{tie}}$ and output $y = \Delta \omega$ data for each LCA recorded for only $10$ seconds at a sampling rate of $0.1$ seconds, which is significantly shorter than the duration of 1200 seconds for ambient data and 600 seconds for LLPRN reported in the literature \cite{zhou2006initial}. 


Regarding the tuning parameters, we used $T = 101$ historical data points for each LCA, collected sequentially with only one LCA being excited at a time. The length of recent past data used in \eqref{Eq:DistEst1-LCA} and \eqref{Eq:RegularizedDDEstimator-LCA} was $T_{\rm ini} = 7$; larger values were found to produce no benefit. The controller gain $\varepsilon$ in \eqref{Eq:DistEst1-LCA} was set via tuning to $\varepsilon = 0.2$ by starting from a small value and increasing until satisfactory performance was reached. For the penalty parameters in \eqref{Eq:RegularizedDDEstimator-LCA}, we set $\lambda_1$ to a large value of $1 \times 10^8$, according to the insights from Section \ref{Sec:RobustDisturbanceEstimator}, and $\lambda_2 = 1 \times 10^2$ was set via tuning by gradually increasing its value until no noticeable improvement in performance was observed.


\subsection{Simulation Scenarios}
\label{Sec:Scenarios}

We consider four different testing scenarios, which aim to highlight the diverse challenges that can arise in a power system, including renewable resource variability, sudden changes in load demand, and equipment failures. The scenarios are
\begin{enumerate}
\item[(1)] response to sudden load changes of different sizes,
\item[(2)] response to solar and wind farm variability, 
\item[(3)] response to a three-phase-to-ground fault, and
\item[(4)] response after loss of a conventional generation unit.
\end{enumerate}


 For all scenarios, we integrate our disturbance estimators into  the hierarchical fast frequency control architecture proposed in \cite{ekomwenrenren2021hierarchical} and compare the model-based disturbance estimator of that work against the data-driven disturbance estimators presented in this paper. We term the controller described in \eqref{Eq:DistEst1-LCA} as the Linear Data-Driven Disturbance Estimator (LDDE) and that in \eqref{Eq:RegularizedDDEstimator-LCA} as the Optimization-Based Data-Driven Disturbance Estimator (ODDE); the ODDE is the default data-driven controller presented in the figures when no other context is given. As a baseline, we compare to the response without any supplementary control scheme, where frequency support is provided only through the primary droop control action of both generators and IBRs. Additionally, we compare against the response obtained by implementing standard automatic generation control (AGC) on the three-area system in Figure \ref{fig:ThreeLCASystem}.\footnote{See \cite[Remark II.4]{ekomwenrenren2021hierarchical} for extensive discussion on the distinctions between the proposed approach and traditional AGC.} In Scenario \#1, we have compared the ODDE against all the alternatives listed above, and demonstrate its performance premium relative to the LDDE. In the remainder of the scenarios, we focus the plots on comparing the better estimator (ODDE) against the model-based approach presented in \cite{ekomwenrenren2021hierarchical}. Finally, the data collection and real-time simulation steps include measurement noise, modelled as zero-mean white noise of standard deviation $10^{-6}$ p.u. for frequency deviation and $2\times 10^{-2}$ p.u. for inter-area power flow measurements; these represent realistic noise on the variables scaled for their typical values (e.g.,  \cite{huang2021decentralized}).

 

\subsection{Scenario \#1: Step Load Changes}
\label{Sec:SelfSuff}
This scenario evaluates the performance of our controller in response to step load changes of varying magnitudes -- 14 MW and 60 MW -- applied in Area 2.

\begin{figure}[ht!]
\centering
\includegraphics[width=1\columnwidth]{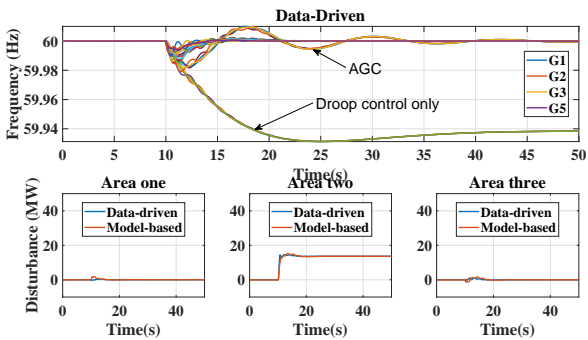}
\caption{Frequency and disturbance estimate during a 14 MW load
change at bus 8 in Area 2.}
\label{fig:SecondSufficientfreqfull_3LCA}
\end{figure}

\begin{figure}[ht!]
\centering
\includegraphics[width=1\columnwidth]{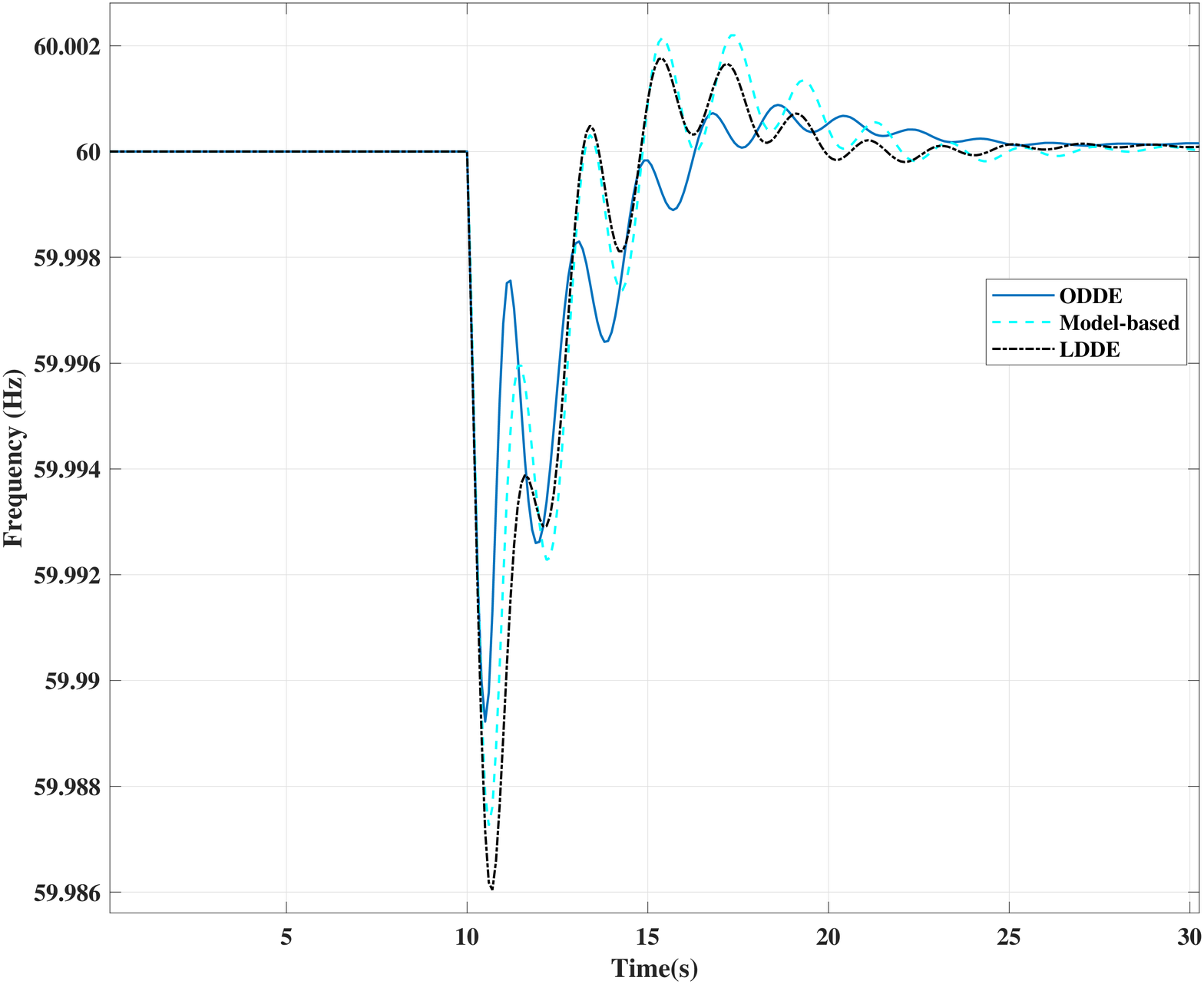}
\caption{Zoomed-in frequency plot of the contingent area showing the control alternatives during a 14 MW load
change at bus 8 in Area 2.}
\label{fig:SecondSufficientfreqzoomin_3LCA}
\end{figure}

\begin{figure}[ht!]
\centering
\includegraphics[width=1\columnwidth]{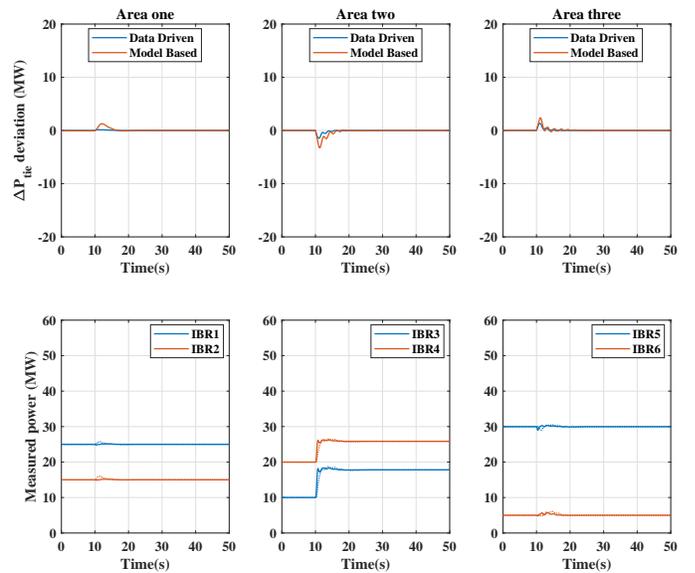}
\setlength{\belowcaptionskip}{-20pt}
\caption{Tie-line deviation and active power profiles during a 14 MW
load change; dashed lines in the lower plots indicate the responses
under model-based estimation.}
\label{fig:SecondSufficientpdev_3LCA}
\end{figure}

At $t=10$s, a small load change of 14 MW is applied at bus 8 in area two. The size of the disturbance is chosen such that the resulting frequency deviation is below the 36 mHz deadband of the generator primary control systems. The frequency response and disturbance estimate of the system are plotted in Figure~\ref{fig:SecondSufficientfreqfull_3LCA}, while the net-tie line deviations and IBR power outputs are displayed in Figure~\ref{fig:SecondSufficientpdev_3LCA}. For clarity in differentiating the alternative approaches, a zoomed-in frequency response plot is shown in Figure~\ref{fig:SecondSufficientfreqzoomin_3LCA}. 


Using both the model-based and data-driven disturbance estimators, the disturbance was quickly identified to originate in Area 2 and promptly corrected by adjusting the setpoints of the local IBRs, with minimal impact on other areas. Overall, the frequency was restored quickly, and the variables in the non-contingent areas returned to their pre-disturbance state due to the decentralized nature of the control scheme. The plots also demonstrate that the proposed optimization-based data-driven estimator outperforms the linear data-driven and model-based estimators in terms of a higher nadir and faster settling time for the post-contingency frequency. We believe the improved performance of the optimization-based estimator relative to the linear estimator is due to its ability to better capture the
dynamics of the true underlying system in terms of behaviour and complexity.

\begin{figure}[ht!]
\centering
\includegraphics[width=1\columnwidth]{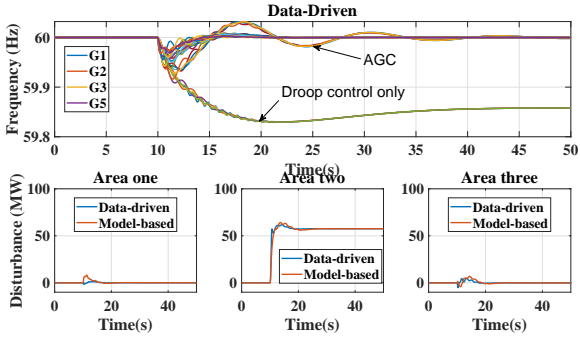}
\caption{Frequency and disturbance estimate during a 60 MW load
change at bus 8 in Area 2.}
\label{fig:FirstSufficientfreqfull_3LCA}
\end{figure}

\begin{figure}[ht!]
\centering
\includegraphics[width=1\columnwidth]{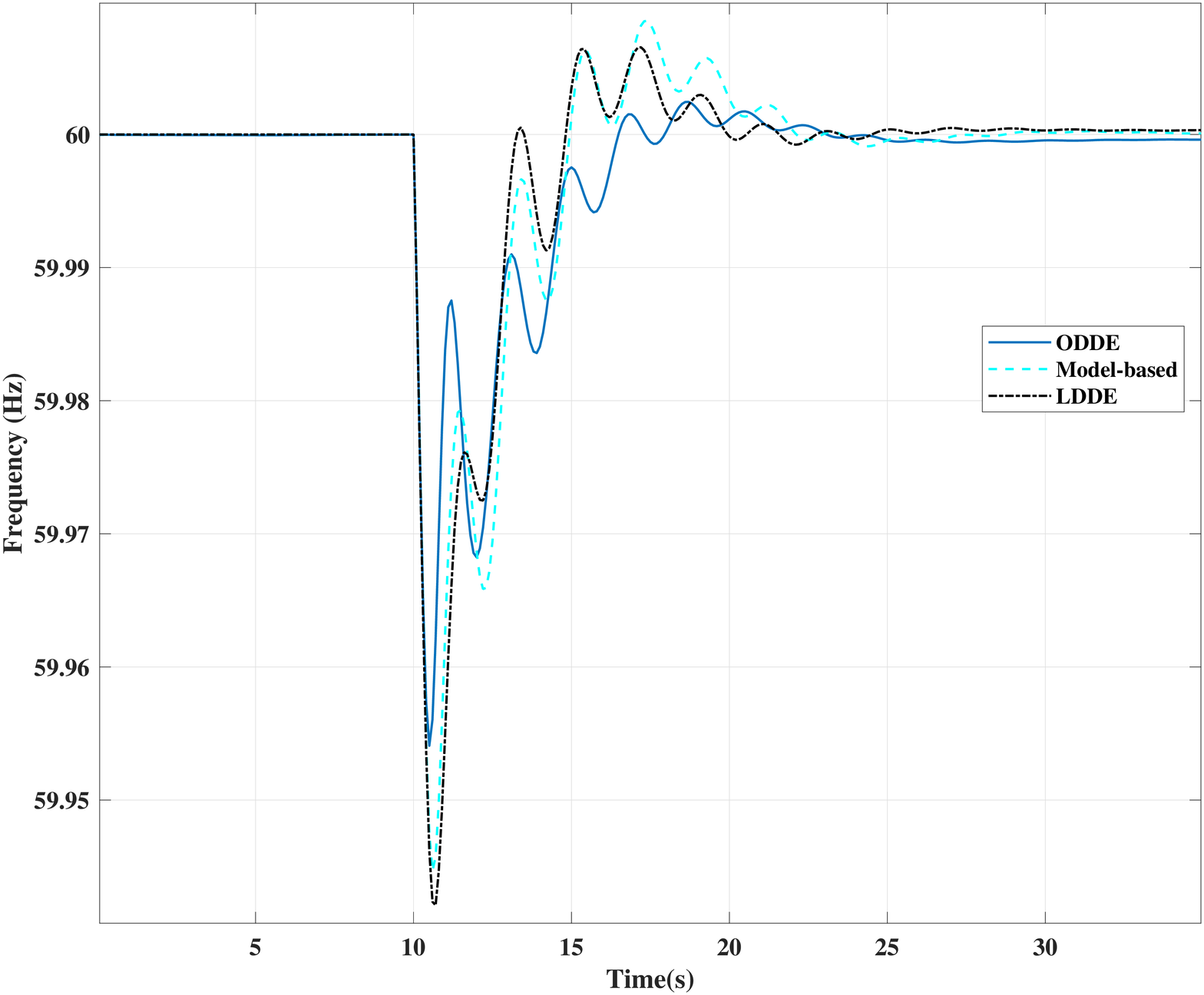}
\caption{Zoomed-in frequency plot of the contingent area showing the control alternatives during a 60 MW load
change at bus 8 in Area 2.}
\label{fig:FirstSufficientfreqzoomin_3LCA}
\end{figure}

\begin{figure}[ht!]
\centering
\includegraphics[width=1\columnwidth]{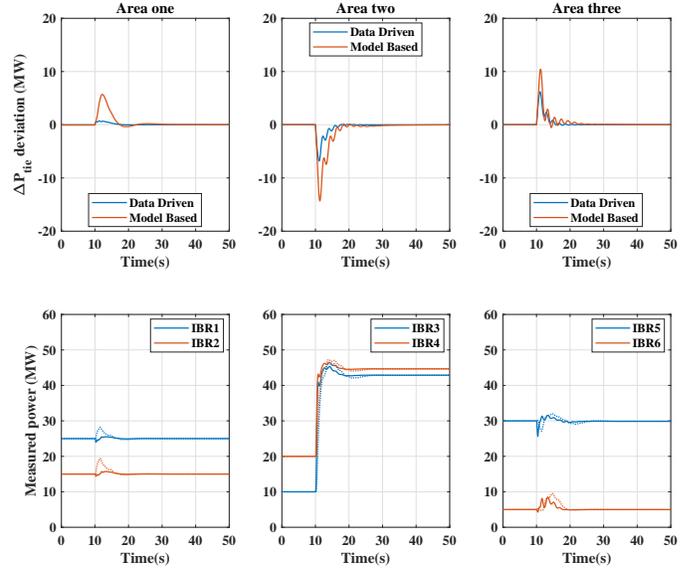}
\setlength{\belowcaptionskip}{-10pt}
\caption{Tie-line deviation and active power profiles during a 60 MW
load change; dashed lines in the lower plots indicate the responses
under model-based estimation.}
\label{fig:FirstSufficientpdev_3LCA}
\end{figure}

The plots in Figures~\ref{fig:FirstSufficientfreqfull_3LCA}, \ref{fig:FirstSufficientfreqzoomin_3LCA}, and \ref{fig:FirstSufficientpdev_3LCA}, which display the frequency response, disturbance estimate, net tie-line deviations, and IBR outputs in response to a step load change of 60 MW applied at bus 8 in Area 2 at $t=10s$, lead to the same conclusion about the controller's performance as the previous scenario with a 14 MW load change. This demonstrates that the proposed controller exhibits superior performance for step load changes both inside and outside the governor deadband range.

In general, the results show that the robust data-driven approach presented in this study outperforms the model-based approach and other alternatives, quickly localizing and compensating for large and small disturbances.




\subsection{Scenario \#2: High Renewable Resource Fluctations}
\label{Sec:Variability}

\begin{figure}[ht!]
\centering
\includegraphics[width=1\columnwidth]{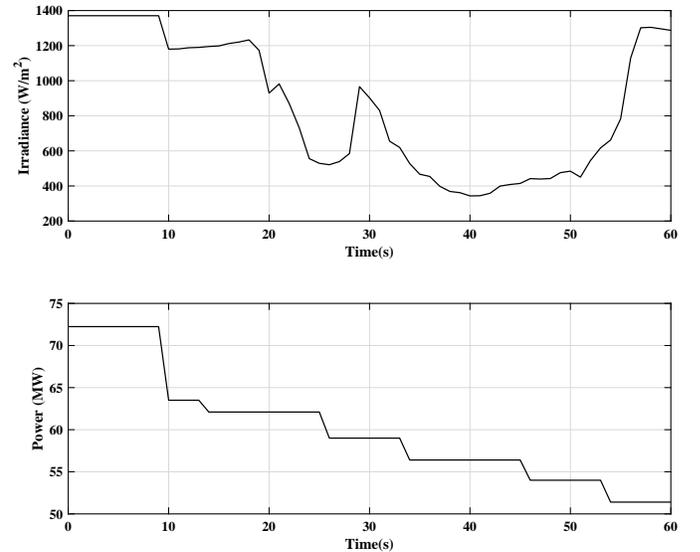}
\caption{Solar irradiance and wind power data representing suitably scaled slices of data on 31 July 2010 and 15 Aug 2019 from \cite{sengupta2010oahu} and \cite{Godahewa2020} repositories.}
\label{fig:RESVolData_3LCA}
\end{figure}

In this scenario, we aim to demonstrate the effectiveness of our data-driven approach in the presence of renewable variability using realistic wind and solar irradiance data. To simulate the solar irradiance component, we use data from the Oahu solar measurement grid 1-year archive \cite{sengupta2010oahu}, containing 1-second measurements of solar irradiance. We select a slice of data from July 31, 2010 (see Figure~\ref{fig:RESVolData_3LCA}). These values are used to simulate a converter-interfaced PV farm in Area 2.

For the wind farm component, we use 4-second resolution wind power measurements from the wind power dataset accessible at \cite{Godahewa2020}, originally obtained from the Australian Energy Market Operator (AEMO). We select a slice of data from August 15, 2019 (see Figure~\ref{fig:RESVolData_3LCA}). These values are used to simulate a converter-interfaced wind farm in Area 1.

\begin{figure}[ht!]
\centering
\includegraphics[width=1\columnwidth]{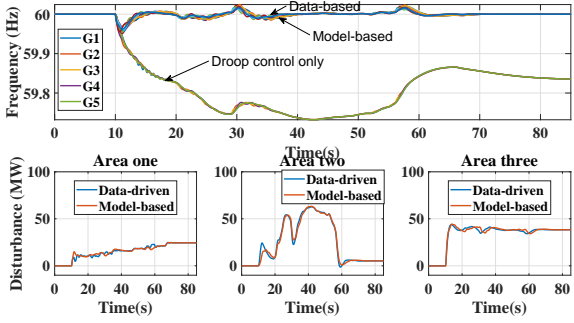}
\setlength{\belowcaptionskip}{-15pt}
\caption{Frequency and disturbance estimate during high renewable resource fluctuations in
multiple areas.}
\label{fig:RESVolfreqfull_3LCA}
\end{figure}

\begin{figure}[ht!]
\centering
\includegraphics[width=1\columnwidth]{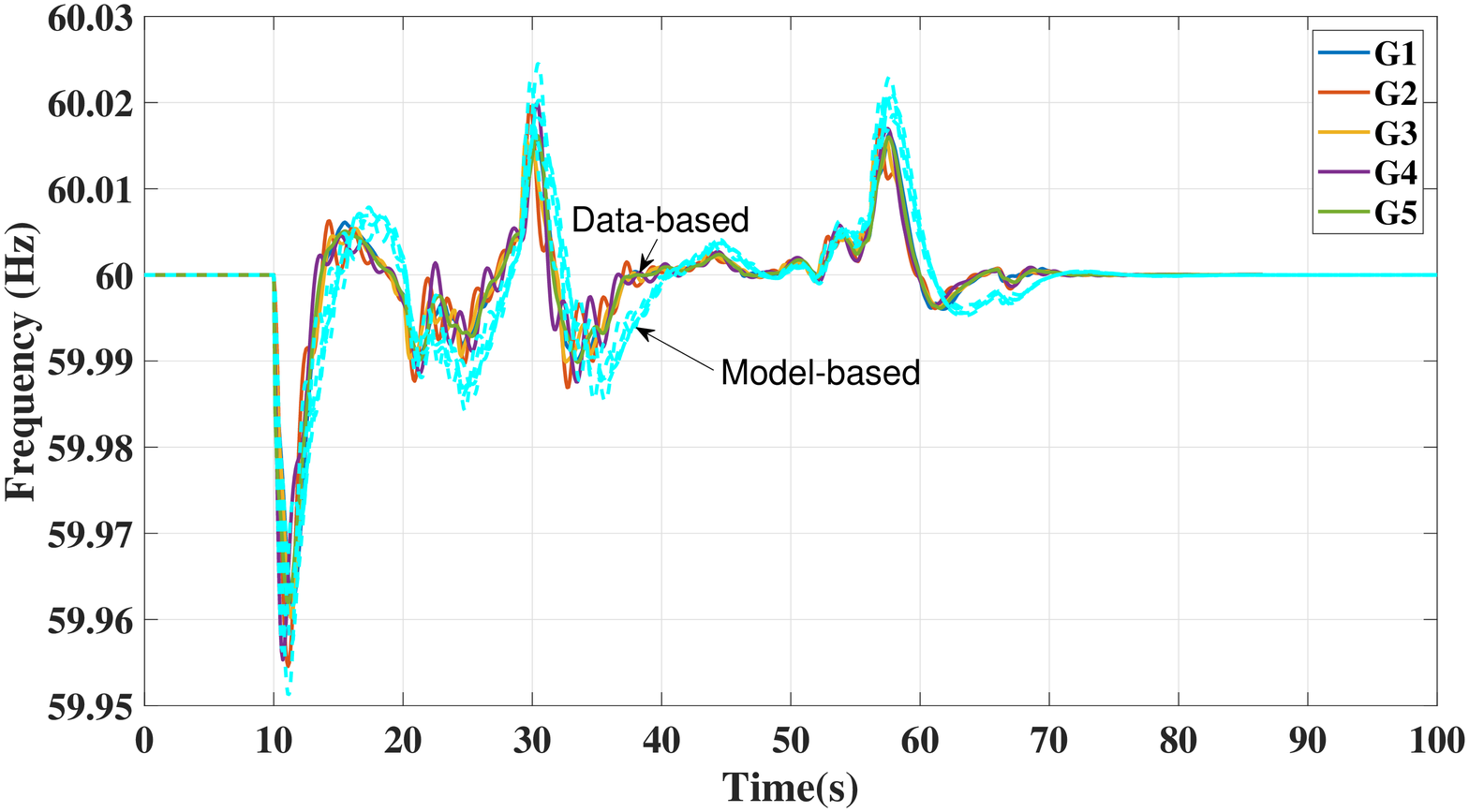}
\setlength{\belowcaptionskip}{-15pt}
\caption{Zoomed-in frequency plot during high renewable resource fluctuations in
multiple areas.}
\label{fig:RESVolfreqzoomin_3LCA}
\end{figure}

\begin{figure}[ht!]
\centering
\includegraphics[width=1\columnwidth]{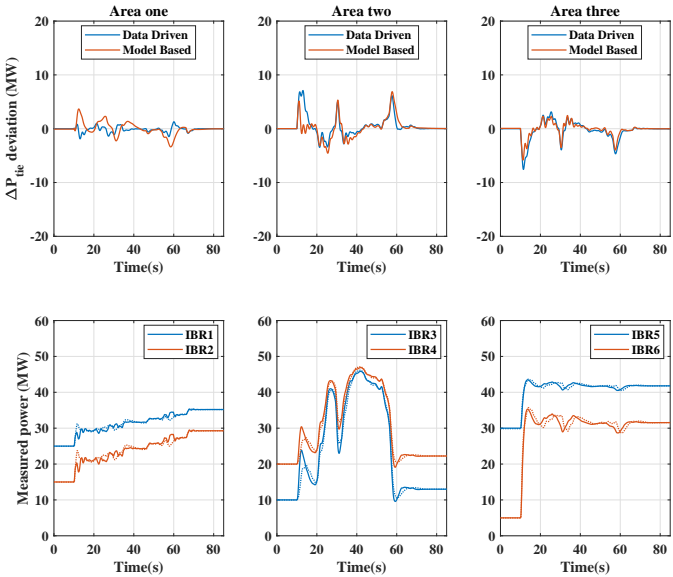}
\caption{Tie-line deviation and active power profiles during high renewable resource fluctuations in
multiple areas.; dashed lines in the lower plots indicate the responses
under model-based estimation.}
\label{fig:RESVolpdev_3LCA}
\end{figure}

At $t=10s$, we introduce a step load change of $40$ MW in Area 2 and simultaneously introduce the new the solar irradiance levels and wind power in accordance with the real-world data. The frequency response, disturbance estimate, net tie-line deviations, and IBR outputs of the power system for this scenario are displayed in Figures~\ref{fig:RESVolfreqfull_3LCA}, \ref{fig:RESVolfreqzoomin_3LCA}, and \ref{fig:RESVolpdev_3LCA}.

The simulation results show that the optimization-based data-driven estimator is able to quickly estimate the highly variable power imbalance in real-time, allowing the frequency to be kept close to the nominal value throughout the simulation period. The performance of our designed data-driven estimator is quantified in Table~\ref{tab:RMSE}, which shows the root mean squared frequency deviation for each generator in the system. The results show that the data-driven disturbance estimator with regularization performs the best.

\begin{table}[ht]
   
\begin{center}
\scalebox{0.8}{
\begin{tabular}{|c|c|c|c|c|c|c|} 
\hline
\textbf{Control type} & \multicolumn{5}{c|}{\textbf{RMSE (Hz)}} & \textbf{Total (Hz)} \\ [0.5ex] 
\hline\hline
 & \textbf{G1} & \textbf{G2} & \textbf{G3} & \textbf{G4} & \textbf{G5} & \\ [0.5ex] 
\hline
Data-based w/ reg. & 0.0065  &  0.0065 &   0.0064  &  0.0066  &  0.0065 & \textbf{0.0325} \\ 
\hline
Data-based w/o reg. & 0.0071  &  0.0073  &  0.0072  &  0.0074  &  0.0073 & 0.0365 \\
\hline
Model-based & 0.0081  &  0.0081  &  0.0080  &  0.0080 &  0.0081 &  0.0403 \\
\hline
AGC & 0.0236 &  0.0225 & 0.0233 & 0.0225 & 0.0233 & 0.1151 \\
\hline
Droop only & 0.1885 &  0.1883 & 0.1884 & 0.1884 & 0.1884 & 0.9420 \\ [1ex] 
\hline
\end{tabular}
}
\end{center}
\setlength{\belowcaptionskip}{-10pt}
\caption{Root Mean Square Error for Control Alternatives.}
\label{tab:RMSE}
\end{table}

\subsection{Scenario \#3: Symmetric Three-Phase Fault}
\label{Sec:SymFault}

\begin{figure}[ht!]
\centering
\includegraphics[width=1\columnwidth]{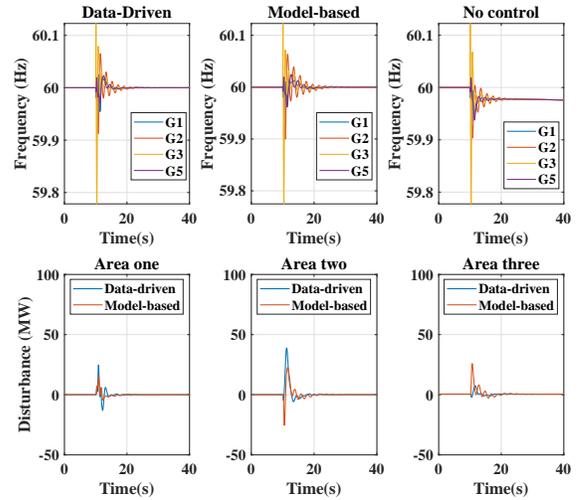}
\caption{Frequency and disturbance estimate during a three-phase fault in Area 2.}
\label{fig:threephasefaultfreqfull_3LCA}
\end{figure}

\begin{figure}[ht!]
\centering
\includegraphics[width=1\columnwidth]{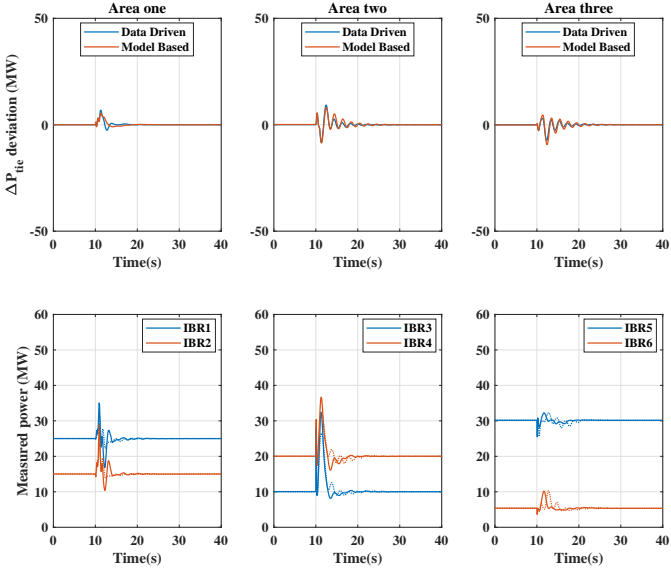}
\caption{Tie-line deviation and active power profiles during a three-
phrase fault in Area 2; dashed lines in the lower plots indicate the
responses under model-based estimation.}
\label{fig:threephasefaultpdev_3LCA}
\end{figure}

The essence of this scenario is to assess the performance of our control approach under a severe contingency like a symmetrical three-phase line-to-ground fault. The response of the system during the fault introduced at bus 8 in Area 2 at $t = 2$s and cleared after $0.1$s is shown in Figures \ref{fig:threephasefaultfreqfull_3LCA} and \ref{fig:threephasefaultpdev_3LCA}. Note that despite the transients, the controller is able to discern that there is no net load imbalance within the area. The results indicate that the controller is able to effectively detect and respond to frequency events, and the data-driven estimator's performance is satisfactory and similar to the model-based estimator.


\subsection{Scenario \#4: Loss of Generator}
\label{Sec:GeneratorLoss}
\begin{figure}[ht!]
\centering
\includegraphics[width=1\columnwidth]{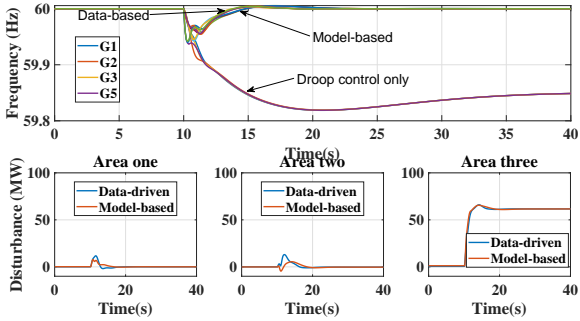}
\setlength{\belowcaptionskip}{-10pt}
\caption{Frequency and disturbance estimate during a generator G4 loss in Area 3.}
\label{fig:Generatorlossfreqfull_3LCA}
\end{figure}

This scenario assesses the performance of our control approach under the loss of generator G2 in area 2 at $t = 10 s$. When the generator G2 is lost, the system experiences a disturbance, and the controllers respond to correct the resulting power imbalance. The response of the system under both the model-based and data-driven control is plotted in Figure~\ref{fig:Generatorlossfreqfull_3LCA}. According to the findings, the data-driven controller outperforms the model-based controller, as demonstrated by its faster frequency settling time and lower overshoot. Importantly, despite the data used to design the LCA controller having been collected while G2 was online, the response indicates that the data-driven controller is effective even in the face of significant changes in power system composition and frequency response dynamics. This illustrates the robustness of the design, and provides flexibility for system operators in deciding how frequently they wish to collect new data to update the controller.

\section{Conclusions}
\label{Sec:Conslusions}

We have proposed and validated through detailed simulations a robust data-driven disturbance estimator that allows us to reliably compute the real-time power imbalance in a highly nonlinear power system, and in the presence of measurement noise. This data-driven estimate has been integrated in the hierarchical frequency control architecture initially proposed in \cite{ekomwenrenren2021hierarchical}, to provide a completely model-free approach to provide fast, localized frequency regulation in the power system. An important direction for future research is further investigation into the design of improved excitation input signals for data collection, and integration of this approach with transmission and distribution coordination schemes.



%

\appendices
\section{Proofs}\label{Append:A}

\begin{pfof}{Theorem \ref{Thm:DDDE}}
Under the stated assumptions of controllability, input data persistency of excitation of order $T_{\rm ini} + 1 + n(\mathscr{B})$, and sufficient initialization length $T_{\rm ini} \geq \ell(\mathscr{B})$, it follows from \cite[Prop. 6]{markovsky2008data} that the output predictor \eqref{Eq:DistEst1-a} produces precisely the same values $\hat{y}(t)$ as the LTI system
\begin{equation}\label{Eq:SSEstimator}
\begin{aligned}
\hat{x}(t+1) &= A\hat{x}(t) + Bu(t) + B_d \hat{d}(t)\\
\hat{y}(t) &= C\hat{x}(t) + Du(t)
\end{aligned}
\end{equation}
where the matrices may be taken to be the same as those in \eqref{Eq:LTI}. The disturbance esimator \eqref{Eq:DistEst1} may therefore be expressed as \eqref{Eq:SSEstimator} with \eqref{Eq:DistEst1-b}, which we rewrite together as
\[
{\small
\begin{aligned}
\begin{bmatrix}
\hat{x}(t+1)\\
\hat{d}(t+1)
\end{bmatrix} &= \underbrace{\begin{bmatrix}A & B_d\\
0 & I_{q}\end{bmatrix}}_{:= \mathcal{A}}\begin{bmatrix}\hat{x}(t) \\ \hat{d}(t)\end{bmatrix} + \begin{bmatrix}B \\ 0\end{bmatrix}u(t) - \underbrace{\begin{bmatrix}0 \\ \varepsilon L\end{bmatrix}}_{:=\varepsilon\mathcal{L}}(\hat{y}(t) - y(t))\\
\hat{y}(t) &= \underbrace{\begin{bmatrix}C & 0\end{bmatrix}}_{:= \mathcal{C}}\begin{bmatrix}\hat{x}(t) \\ \hat{d}(t)\end{bmatrix} + Du(t)
\end{aligned}
}
\]
where $y(t)$ is the measured output of \eqref{Eq:LTI}. The above has the form of a Luenberger observer, and standard estimation error analysis (e.g., \cite{chen1984linear}) implies that we will have $\hat{d}(t) \to d(t)$ exponentially, and irrespective of the initial conditions, if 
\[
\mathcal{A} - \varepsilon\mathcal{L}C = \begin{bmatrix}A & B_d\\
-\varepsilon LC & I_q
\end{bmatrix}
\]
is Schur stable. Recall that, by assumption, $A$ is Schur stable, so $I_n-A$ is invertible; based on this define the invertible matrix
\[
T = \begin{bmatrix}I_n & (I_n-A)^{-1}B_d\\
0 & I_q\end{bmatrix}.
\]
By similarity, $\mathcal{A} - \varepsilon\mathcal{L}C$ is Schur stable if and only if $\mathcal{M}(\epsilon) := T(\mathcal{A} - \varepsilon\mathcal{L}C)T^{-1}$ is as well. Simple calculations show that $\mathcal{M}(\epsilon)$ evaluates to
\[
\mathcal{M}(\epsilon) = \begin{bmatrix}
A + \varepsilon M_1 & \varepsilon M_2\\
\varepsilon M_3 & I_q - \varepsilon LG(1)
\end{bmatrix},
\]
where $M_1,M_2,M_3$ are constant matrices and $G(1) = C(I_n-A)^{-1}B_d$. By assumption, $G(1)$ has full column rank and $L = G(1)^{\dagger}$; thus, we have that $LG(1) = I_q$, and the $(2,2)$ block of the above  simplifies to $(1-\varepsilon)I_q$. Since $A$ is Schur stable, by standard linear Lyapunov theory there exists a matrix $P \succ 0$ such that $A^{\sf T}PA - P \prec 0$. Defining $\mathcal{P} = \mathrm{diag}(P,I_q)$, straightforward calculations and a use of the Schur complement lemma show that $\mathcal{M}(\epsilon)^{\sf T}\mathcal{P}\mathcal{M}(\epsilon) - \mathcal{P} \prec 0$ for all sufficiently small $\varepsilon > 0$, which establishes that $\mathcal{M}(\epsilon)$ is Schur stable and completes the proof.
\end{pfof}

\ifCLASSOPTIONcaptionsoff
  \newpage
\fi



%
\bibliographystyle{ieeetr}
\bibliography{brevalias,Main,JWSP,paper,comments}

\end{document}